\documentclass[prl,aps,twocolumn,showpacs,longbibliography, superscriptaddress]{revtex4-1}
\usepackage{graphicx}
\usepackage{amsmath}
\usepackage{bm}
\usepackage{amstext}
\usepackage{amsxtra}
\usepackage{float}
\usepackage{upgreek}

\usepackage{times}

\begin{document}

\title{Universal Scaling Laws in the Dynamics of a Homogeneous Unitary Bose Gas}

\author{Christoph Eigen}
\affiliation{Cavendish Laboratory, University of Cambridge, J. J. Thomson Avenue, Cambridge CB3 0HE, United Kingdom }
\author{Jake A. P.  Glidden}
\affiliation{Cavendish Laboratory, University of Cambridge, J. J. Thomson Avenue, Cambridge CB3 0HE, United Kingdom }
\author{Raphael Lopes}
\affiliation{Cavendish Laboratory, University of Cambridge, J. J. Thomson Avenue, Cambridge CB3 0HE, United Kingdom }
\author{Nir Navon}
\affiliation{Cavendish Laboratory, University of Cambridge, J. J. Thomson Avenue, Cambridge CB3 0HE, United Kingdom }
\affiliation{Department of Physics, Yale University, New Haven, CT 06511, USA}
\author{Zoran Hadzibabic}
\affiliation{Cavendish Laboratory, University of Cambridge, J. J. Thomson Avenue, Cambridge CB3 0HE, United Kingdom }
\author{Robert P. Smith}
\email{rps24@cam.ac.uk}
\affiliation{Cavendish Laboratory, University of Cambridge, J. J. Thomson Avenue, Cambridge CB3 0HE, United Kingdom }

\begin{abstract}

We study the dynamics of an initially degenerate homogeneous Bose gas after an interaction quench to the unitary regime at a magnetic Feshbach resonance.
As the cloud decays and heats, it exhibits a crossover from degenerate- to thermal-gas behaviour, both of which are characterised by universal scaling laws linking the particle-loss rate to the total atom number $N$. In the degenerate and thermal regimes the per-particle loss rate is $\propto N^{2/3}$ and $N^{26/9}$, respectively. The crossover occurs at a universal kinetic energy per particle and at a universal time after the quench, in units of energy and time set by the gas density.
By slowly sweeping the magnetic field away from the resonance and creating a mixture of atoms and molecules, we also map out the dynamics of correlations in the unitary gas, which display a universal temporal scaling with the gas density, and reach a steady state while the gas is still degenerate.

\end{abstract}

\date{\today}



\maketitle


Strong interactions and correlations are at the heart of the most interesting many-body quantum phenomena. The possibility to control the interaction strength via Feshbach resonances~\cite{Chin:2010} makes ultracold atomic gases an excellent setting for studies of strongly correlated behaviour.  On resonance, the $s$-wave scattering length $a$, which characterises two-body contact interactions, diverges. In this so-called unitary regime the interactions are as strong as allowed by quantum mechanics, and  the physics cannot explicitly depend on $a$,  leading to the possibility of new types of universal behaviour~\cite{Ho:2004a, Inguscio:2007, Zwerger:2011,Zwierlein:2014, Chevy:2016}.

Of particular interest are the interaction-dominated degenerate unitary gases. Within the `universality hypothesis', they have only one relevant lengthscale - the average interparticle spacing, given by the density $n$, which also sets the natural energy and time scales~\cite{Ho:2004a}:
\begin{equation}
E_n=\frac{\hbar^2}{2m} (6\pi^2 n)^{2/3}
\quad\text{and}\quad
t_n=\hbar/E_n \, ,
\label{eq:Entn}
\end{equation}
where $m$ is the atom mass. These `Fermi energy' and `Fermi time' scales are applicable to both Fermi and Bose gases.
In Bose gases, however, the universality can be broken by Efimov physics~\cite{Efimov:1970,Kraemer:2006, Braaten:2007, Zaccanti:2009, Ferlaino:2011, Wild:2012,Machtey:2012,Roy:2013, Fletcher:2017,Klauss:2017,DIncao:2017}. The Feshbach dimer molecular state, responsible for the resonance, is of size $a$ and becomes unbound as $a \rightarrow \infty$, but the infinite series of Efimov trimer states, each of a size 22.7 times larger than the previous one, can introduce new lengthscales into the problem.

While unitary Fermi gases have been extensively explored \cite{Inguscio:2007, Zwerger:2011,Zwierlein:2014}, the experimental~\cite{Rem:2013, Fletcher:2013, Makotyn:2014, Eismann:2016, Fletcher:2017, Klauss:2017} and theoretical~\cite{Diederix:2011, Li:2012, Yin:2013, Sykes:2014, Jiang:2014, Rancon:2014, Laurent:2014, Rossi:2014, Smith:2014, Piatecki:2014, Jiang:2016, Comparin:2016, Yin:2016} studies of unitary Bose gases are only recently emerging.
An experimental challenge is that they exhibit rapid three-body loss and heating,
which also raises fundamental questions about the extent to which they have well defined equilibrium properties~\cite{Makotyn:2014}, but the loss dynamics also offer a valuable probe of the unitary behaviour~\cite{Rem:2013, Fletcher:2013, Makotyn:2014, Eismann:2016,Klauss:2017}.  While coherent three-body correlations~\cite{Fletcher:2017} and Efimov trimers~\cite{Klauss:2017} have been observed, the decay dynamics~\cite{Rem:2013, Fletcher:2013,  Makotyn:2014, Eismann:2016, Klauss:2017} have been consistent with universal scalings (see also~
\cite{UnitaryFootnote1}).
%
All experiments so far were performed with harmonically trapped gases, and their interpretation relies on knowledge of the inhomogeneous density profiles. For a degenerate gas, the density profile is known prior to a quench to unitarity, and hence right after it, but the subsequent evolution is complicated by strong interactions and inhomogeneous losses and heating.


In this Letter, we study the dynamics of a {\it homogeneous}, initially degenerate Bose gas quenched to unitarity. In our gas, produced in an optical box trap~\cite{Gaunt:2013}, $E_n$ and $t_n$ are global variables simply set by the total atom number $N$. We can thus quantitatively study the full evolution of a cloud as it decays and heats.
In both degenerate and thermal regimes we observe (different) universal atom-loss scaling laws in agreement with theoretical predictions, and we characterise the universal features of the crossover between the two regimes. By slowly ramping the magnetic field away from the resonance and creating an atom-molecule mixture, we also study the dynamics of correlations in the unitary gas. These correlations show $t_{n_0}$ (where $n_0$ is the initial density) as the only relevant timescale, and indicate that the system reaches a strongly correlated quasi-equilibrium state while it is still degenerate.

\begin{figure}[t!]
\centering
\includegraphics[width=1\columnwidth]{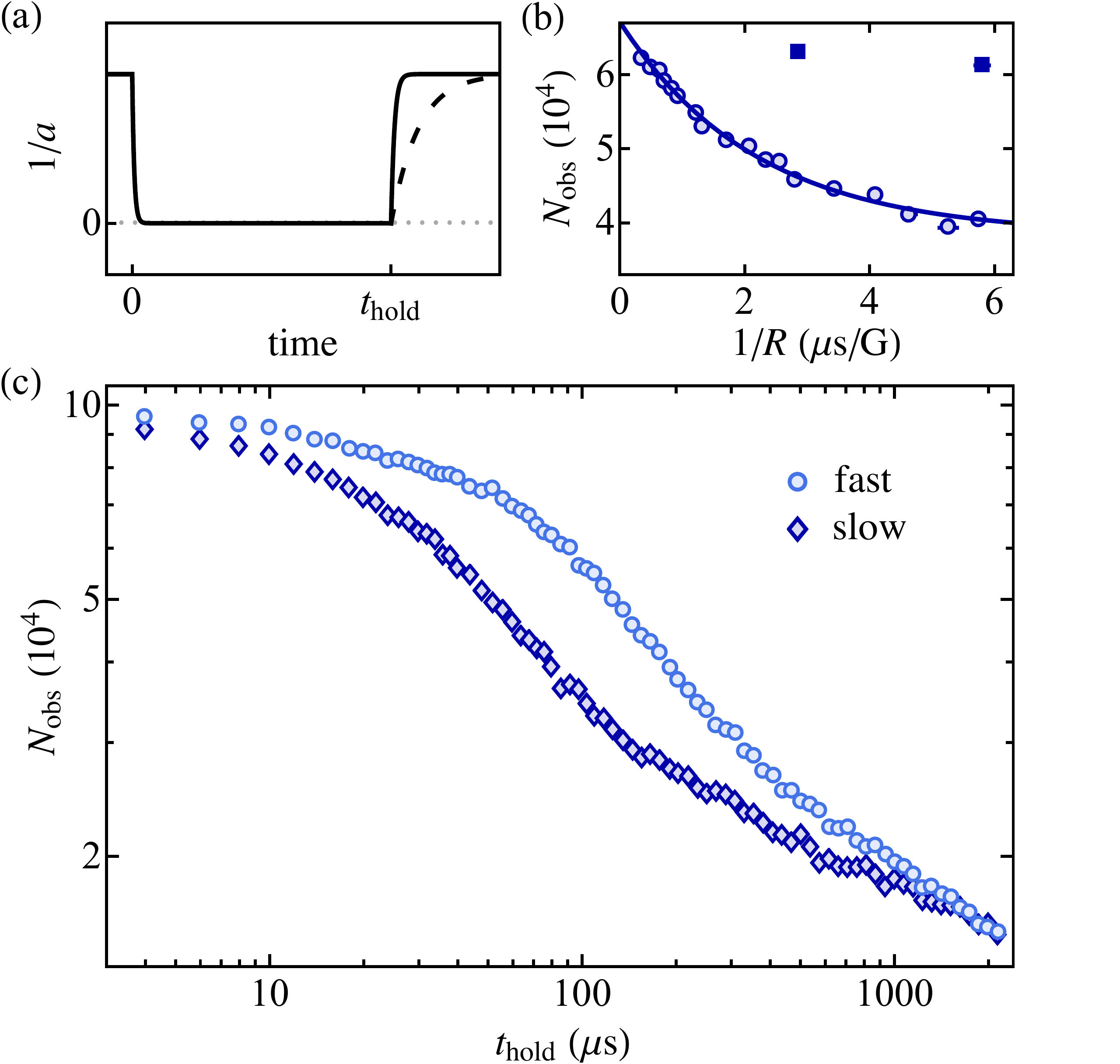}
\caption{
Measurement of atom-loss and correlation dynamics.
(a) A BEC is quenched to unitarity, the cloud is held there for a variable time $t_{\rm hold}$, and then ramped away from unitarity at a variable rate $R = - {\rm d} B/ {\rm d} t$. An infinitely fast ramp-out would project the resonantly interacting cloud onto free-atom states, while a finite-rate ramp-out creates a mixture of atoms and molecules.
(b) Open symbols: The observed atom number versus $1/R$, for $N_0 = 98 \times 10^3$ and $t_{\rm hold}=80~\mu$s.  Solid line is an exponential fit, which gives an exponential constant of $2.2(3)~\mu$s/G. Solid symbols: $N_{\rm obs}$ if the molecules are dissociated by a second magnetic-field pulse to resonance.
(c) Evolution of $N_{\rm obs}$ with $t_{\rm hold}$ for
our fastest ramp-out ($0.3~\mu$s/G) and a much slower one ($6~\mu$s/G).
The fast ramp-out data shows the on-resonance atom loss, and the difference between the two curves reveals the correlation dynamics in the unitary gas.
}
\label{Fig1}
\end{figure}

Our clouds are produced in the lowest hyperfine state of $^{39}$K, which has a background scattering length of $- 29\,a_0$ (where $a_0$ is the Bohr radius) and a 52-G wide Feshbach resonance centred at $B_0 =402.70(3)$~G~\cite{DErrico:2007, Fletcher:2017}. The atoms are confined in a cylindrical box trap of radius 15(1)~$\mu$m and length 50(2)~$\mu$m~\cite{Gaunt:2013, Eigen:2016}, and we vary the initial atom number $N_0$ in the range $(48 - 214)\times 10^3$. This corresponds to $t_{n_0}$ values of a few tens of microseconds, and the range of fields near $B_0$ where $n_0 a^3 >1$ is $\Delta B \sim 0.1$~G.

We start with a weakly interacting quasi-pure Bose--Einstein condensate (BEC) more than $10 \, \Delta B$ away from resonance, where
$n_0 a^3<10^{-3}$. As outlined in Fig.~\ref{Fig1}(a), we then rapidly (within 2~$\mu$s) quench the magnetic field to $B_0$~
\cite{UnitaryFootnote2},
wait for a time $t_{\rm hold}$, and then ramp the field away from unitarity. We finally image the atoms after $8-32$~ms of time-of-flight (ToF) expansion.
The number of atoms that we observe, $N_{\rm obs}$, can reduce with $t_{\rm hold}$ for two reasons: (i) due to losses at unitarity, and (ii) because the ramp-out from unitarity results in a mixture of atoms and (Feshbach and/or Efimov) molecules~\cite{Klauss:2017}, and our imaging detects only free atoms. The molecular fraction after the ramp-out depends on both the correlations in the unitary gas and the ramp-out rate~\cite{Klauss:2017, Regal:2003b, Altman:2005, Hodby:2005}, $R = - {\rm d} B/ {\rm d} t$~
\cite{UnitaryFootnote3}.
Independently of the many-body state at unitarity, an infinitely rapid ramp-out should essentially project the resonantly interacting cloud onto free-atom states, so that $N_{\rm obs} = N$.

To disentangle  the two sources of the reduction of $N_{\rm obs}$ we vary the ramp-out rate, as illustrated in Fig.~\ref{Fig1}(b). Here the open symbols show $N_{\rm obs}$ versus $1/R$ for $N_0 = 98 \times 10^3$ and a fixed $t_{\rm hold}=80~\mu{\rm s}$.
The data are fitted well by an exponential, characteristic of a Landau-Zener process~\cite{Regal:2003b, Hodby:2005}.
Extrapolating to $1/R=0$, we assess that with our technically-limited fastest ramp-out, $1/R=0.3~\mu$s/G, we get $N_{\rm obs} = N$ to within $<10\%$; we verified this for our full range of $N_0$ and at several $t_{\rm hold}$.
For slower ramp-outs $N_{\rm obs}$ is reduced by up to $40\%$ (for this $N_0$ and $t_{\rm hold}$).
To verify that this reduction occurs due to the creation of an atom-molecule mixture, after the first trip to the resonance (with a slow ramp-out) we apply a brief ($8~\mu$s) second magnetic-field pulse to $B_0$ to break up the molecules, and find that most of the missing atoms reappear - see solid symbols in Fig.~\ref{Fig1}(b)~
\cite{UnitaryFootnote4}.

In Fig.~\ref{Fig1}(c) we show, for $N_0 = 98 \times 10^3$, the evolution of $N_{\rm obs}$ with $t_{\rm hold}$ for both our fastest ramp-out and a much slower one, $1/R=6~\mu$s/G.
Note that there is very little difference between the two curves for very short $t_{\rm hold}$, showing that it takes time for the system to develop the correlations that lead to the creation of an atom-molecule mixture.

\begin{figure}[t!]
\centering
\includegraphics[width=1\columnwidth]{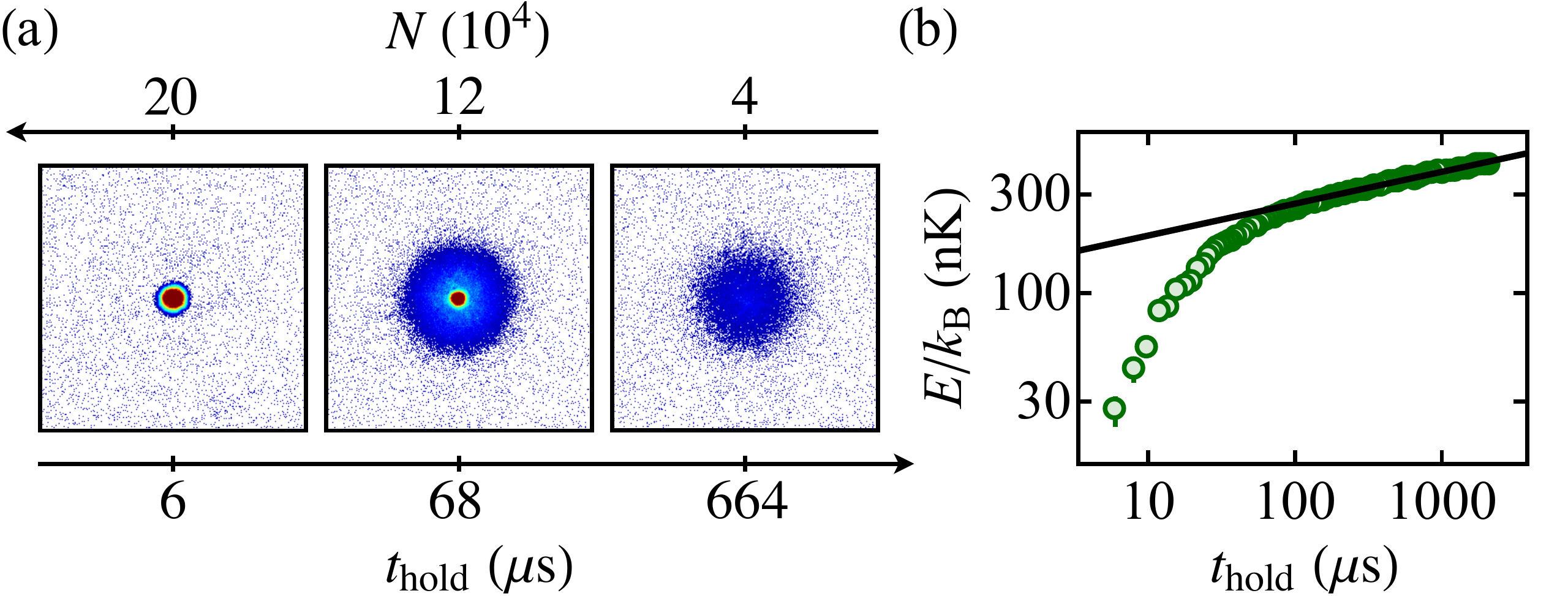}
\caption{
Growth of the kinetic energy, for $N_0=214 \times 10^3$.
(a) Absorption images taken for various $t_{\rm hold}$ and after 12~ms of ToF expansion at weak interactions.
(b) Kinetic energy per particle, $E$, versus $t_{\rm hold}$. Note that the interaction energy per particle after the ramp-out is $< 20$~nK. The solid line shows the $2/13$ power law
predicted for a thermal gas at long $t_{\rm hold}$, where $E \propto T$. }
\label{Fig2a}
\end{figure}

From here on we separately study the atom-loss and correlation dynamics.
We first focus on the atom loss, and assume that for our largest $R$ to a good approximation $N_{\rm obs} = N$.
From the ToF images (see Fig.~\ref{Fig2a}) we also extract the kinetic energy per particle, $E$, which monotonically grows with $t_{\rm hold}$.

The per-particle loss rate is $\dot{N}/N=-L_3 \langle n^2 \rangle $ where $L_3$ is the three-body loss coefficient.
For a homogeneous gas we simply have  $\langle n^2 \rangle = n^2 = N^2/V^2$ (where $V$ is the volume),
and one can generally predict a scaling-law behaviour:
\begin{equation}\label{eq:lossrate}
   {\dot{N}}/{N} = - \, {\rm const}  \times N^{\gamma} \, .
\end{equation}

Away from unitarity $L_3 \propto a^4$~\cite{Fedichev:1996b, Weber:2003}, and in a degenerate unitary gas (assuming universal behaviour) $a$ is replaced by $n^{-1/3} \propto N^{-1/3}$, so $\gamma =2/3$; that is, the instantaneous $t_n$ is the only timescale, so at all times $\dot{N}/N \propto - 1/t_n$.
Such scaling was recently seen in the initial loss rate in a harmonically trapped $^{85}$Rb gas quenched to unitarity~\cite{Klauss:2017}.

In a thermal unitary Bose gas $\dot{N}/N \propto - \langle n^2 \rangle /T^2$~\cite{Rem:2013, Fletcher:2013, Eismann:2016}, where $T$ is the temperature, so in our homogeneous case $\dot{N}/N \propto - N^2 /T^2$. Here $a$ is replaced by the thermal wavelength $\lambda \propto 1/\sqrt{T}$, which is a statistical measure of the inverse relative atomic momenta. We now have two lengthscales, $n^{-1/3}$ and $\lambda$, but their dynamics are coupled and we can still predict a simple scaling law in the form of Eq.~(\ref{eq:lossrate}). Heating occurs because atoms with lower relative momenta have a higher unitarity-limited loss rate and are preferentially lost from the system.
Following~\cite{Rem:2013}, for a homogeneous gas we get ${\rm d}T/T= - (4/9) \, {\rm d}N/N$, so $T \propto N^{-4/9}$, and we thus predict $\gamma=2 + 2 \times 4/9=26/9$. Note that this theory also implies that at long times $T \propto t_{\rm hold}^{2/13}$, in agreement with our measurements shown in Fig.~\ref{Fig2a}(b).

\begin{figure}[t!]
\centering
\includegraphics[width=1\columnwidth]{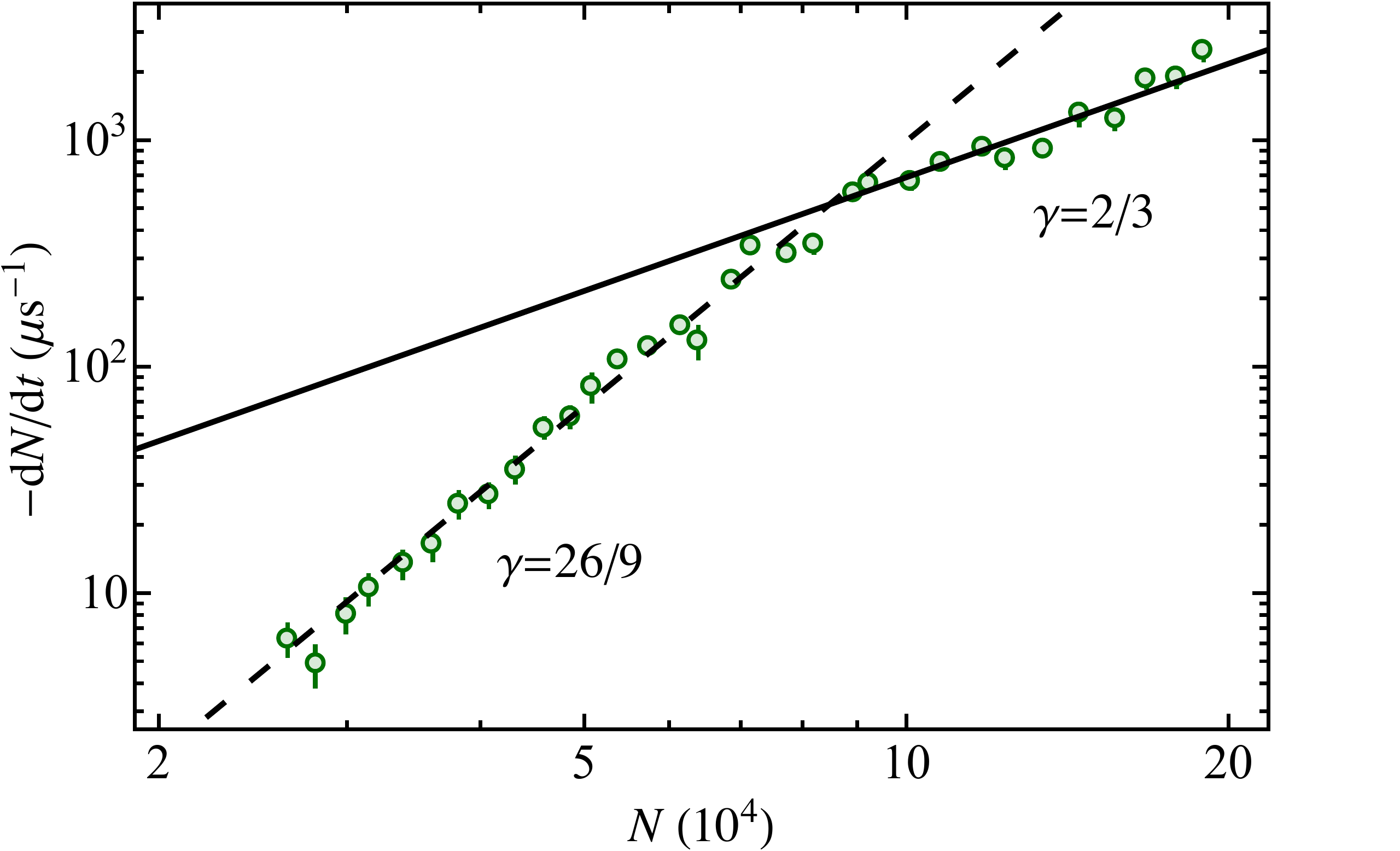}
\caption{
Atom-loss scaling laws, for $N_0=214 \times 10^3$.
We observe both the $\gamma=2/3$ and the $\gamma=26/9$ law, predicted (respectively) for a degenerate and a thermal unitary Bose gas.
The crossover between the two regimes occurs at a well defined atom number.}
\label{Fig2b}
\end{figure}

To experimentally study the instantaneous loss rate, we numerically differentiate our atom-loss curves, $N(t_{\rm hold})$.
In Fig.~\ref{Fig2b} we plot $\dot{N}$ versus $N$ for $N_0=214 \times 10^3$.  We clearly observe both the degenerate-gas $\gamma =2/3$ behaviour, for large $N$ (short $t_{\rm hold}$), and the thermal-gas $\gamma = 26/9$ behaviour, for small $N$ (long $t_{\rm hold}$).

\begin{figure}[t!]
\centering
\includegraphics[width=1\columnwidth]{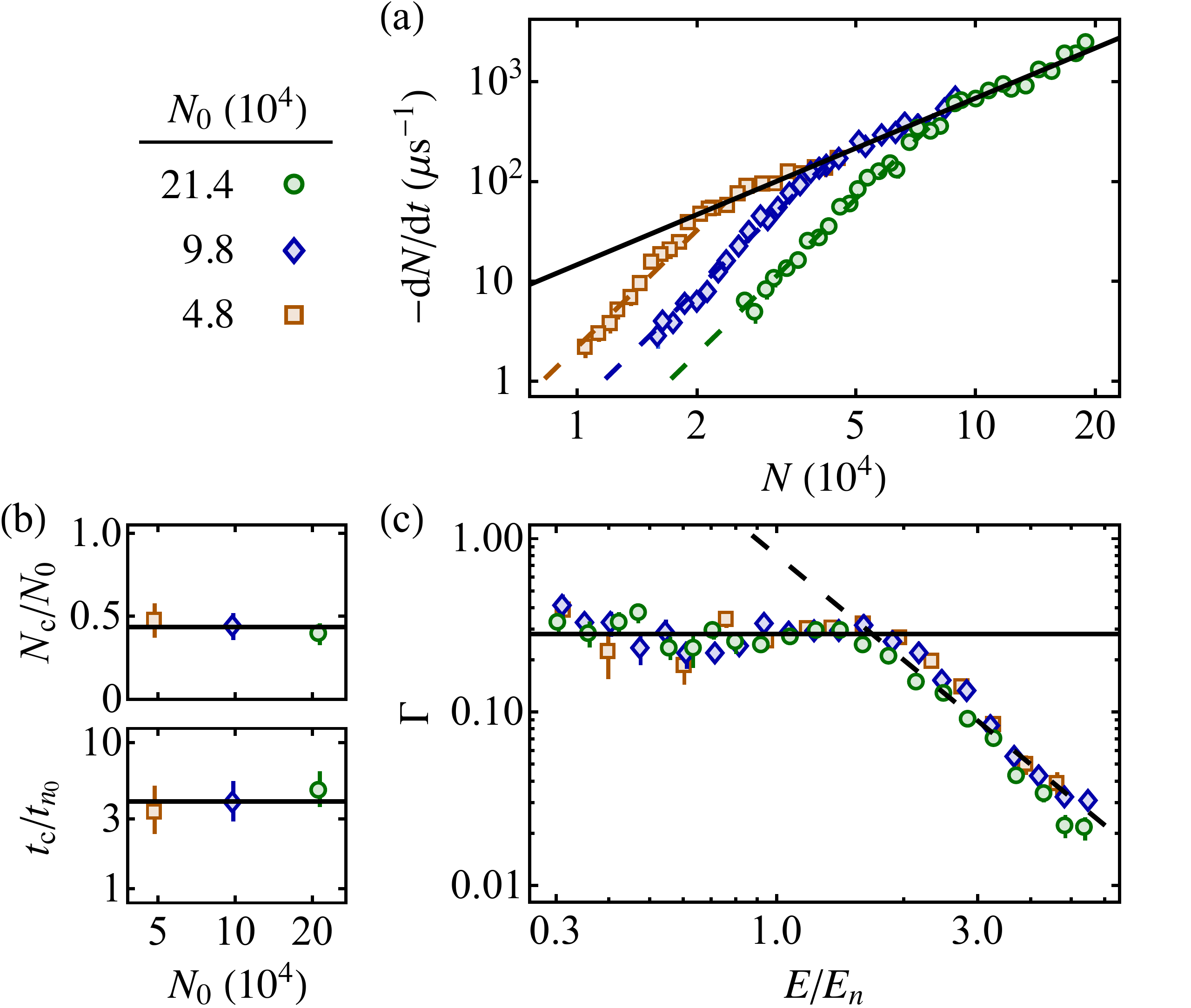}
\caption{Universal crossover from a degenerate to a thermal unitary gas.
(a) The same $2/3$ law, $\dot{N}/N = - 0.28/t_n$ (solid line), describes degenerate gases with different $N_0$.
The dashed lines show the $\gamma=26/9$ scaling.
(b) For all $N_0$ the crossover occurs at almost the same $N/N_0$ and $t_{\rm hold}/t_{n_0}$; averaging gives $N_{\rm c} = 0.43(4) \, N_0$ and $t_{\rm c}=4.0(4)\,t_{n_0}$.
(c) Plotting $\Gamma$ versus $E/E_n$ yields a single universal curve,
with the crossover at $E_{\rm c} = 1.7(2) E_n$. The solid line is $\Gamma = 0.28$ and the dashed one shows the expected $\Gamma \propto (E_n/E)^2$.}
\label{Fig3}
\end{figure}

In Fig.~\ref{Fig3}(a) we plot $\dot{N}$ versus $N$ for different $N_0$ and see that all the degenerate-gas data (large $N/N_0$) follow the same $\gamma=2/3$ law (solid line).
Writing $\dot{N}/N = - A/t_n$, we get $A=0.28(3)$; we assess that due to the $< 10\%$ difference between $N_{\rm obs}$ and the true $N$ we might overestimate $A$ by up to $0.04$.
For comparison, from the $^{85}$Rb data~\cite{Klauss:2017} we extract a slightly lower $A\approx 0.18$, which
is consistent~\cite{Sykes:2014} with the difference in the Efimov width parameters,
$\eta^* \approx 0.09$ for $^{39}$K~\cite{Fletcher:2013} and $\approx 0.06$ for $^{85}$Rb~\cite{Wild:2012}.

For all $N_0$, the small $N/N_0$ data agree with $\gamma = 26/9$ [dashed lines in Fig.~\ref{Fig3}(a)]. Within errors, the crossover to this regime always occurs at the same fraction of the initial population and the same $t_{\rm hold}$ expressed in units of $t_{n_0}$; see Fig.~\ref{Fig3}(b). We get $N_{\rm c}=0.43(4)\,N_0$ and $t_{\rm c}=4.0(4)\,t_{n_0}$~
\cite{UnitaryFootnote5}.

In Fig.~\ref{Fig3}(c) we relate the change in $\gamma$ to the growth of the dimensionless $E/E_n$.
We define the dimensionless per-particle loss rate
\begin{equation}
\Gamma = - t_n \dot{N}/N \, ,
\end{equation}
such that in a degenerate gas $\Gamma = A$ and in a thermal gas $\Gamma \propto N^{4/3}/T^2 \propto (E_n/E)^2$ (using $\dot{N}/N \propto - N^2/T^2$ and $E\propto T$).
Plotting $\Gamma$ versus $E/E_n$, the data for different $N_0$ collapse onto a single universal curve, with the crossover at $E_{\rm c} = 1.7(2) \, E_n$.
For an ideal Bose gas in equilibrium this energy would be quite high, corresponding to $T\approx 3 \, T_{\rm c}$, where $T_{\rm c}$ is the BEC critical temperature.
However, in a unitary gas we expect $E/E_n$ to be of order unity even at $T=0$~\cite{Diederix:2011,Rossi:2014,Jiang:2014}.
An important challenge for future work is to disentangle the contributions to the initial growth of $E/E_n$ (at $t_{\rm hold} < t_{\rm c}$) due to heating and due to the development of the interaction-induced correlations that coherently broaden the momentum distribution~\cite{Makotyn:2014, Diederix:2011, Smith:2014, Comparin:2016, Yin:2016}.


\begin{figure}[t!]
\centering
\includegraphics[width=1\columnwidth]{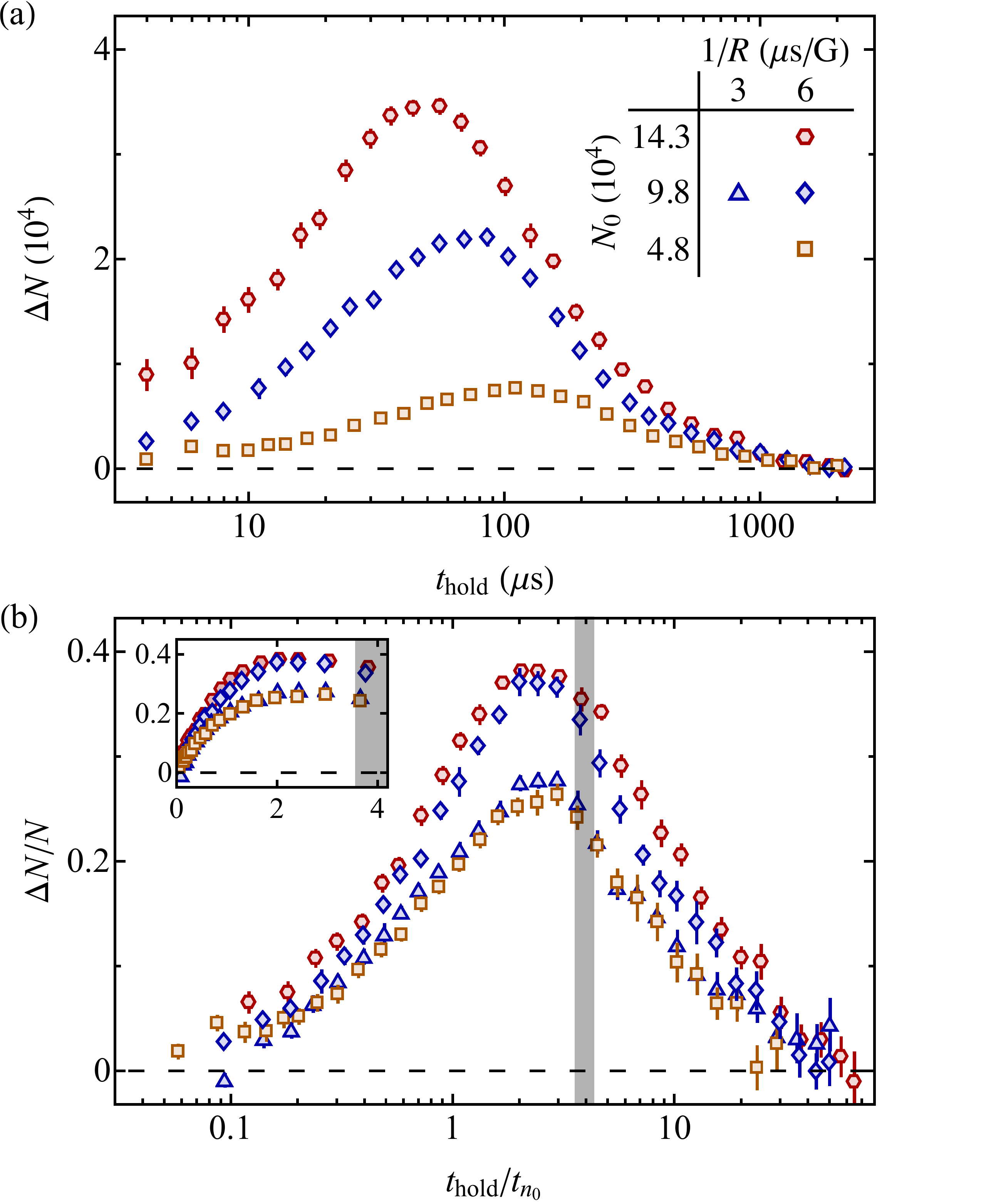}
\caption{Correlation dynamics. (a) $\Delta N$ versus $t_{\rm hold}$ for different $N_0$ and the same slow ramp-out, $1/R = 6~\mu$s/G.
(b) $\Delta N/N$ versus the reduced time $t_{\rm hold}/t_{n_0}$, for different $N_0$ and $1/R$; see legend in (a). The shaded region marks the crossover to the thermal regime at $t_{\rm c}=4.0(4)\,t_{n_0}$. Inset: zoom-in on the degenerate regime shows that the correlations reach a (quasi-)steady state well before $t_{\rm c}$.}
\label{Fig4}
\end{figure}

We now turn to the dynamics of correlations at unitarity. Here we denote by $\Delta N$ the reduction in $N_{\rm obs}$ due to a slow ramp-out (still setting $N_{\rm obs} = N$ for our largest $R$). In Fig.~\ref{Fig4}(a) we show $\Delta N$ versus $t_{\rm hold}$ for different $N_0$ and a same slow ramp-out.
In Fig.~\ref{Fig4}(b) we plot the fractional $\Delta N/N$ versus the dimensionless $t_{\rm hold}/t_{n_0}$, and also show a measurement with a different $R$.
Plotted this way, apart from their heights all the curves look essentially the same, showing that the correlation dynamics are universally determined by the initial density.

Qualitatively, the non-monotonic behaviour of $\Delta N/N$
arises due to the competition of two effects. It takes time for the correlations to develop after the interaction quench, but on the other hand at very long times the system is again uncorrelated because it heats up and the phase space density drops significantly (see also~\cite{Hodby:2005,Werner:2012}).

In the inset of Fig.~\ref{Fig4}(b) we highlight (on a linear time scale) that $\Delta N/N$ becomes essentially flat well before $t_{\rm c}$, meaning that at least in some sense the system reaches a quasi-equilibrium while  it is still degenerate.
The same conclusion was drawn in~\cite{Makotyn:2014} based on momentum-distribution dynamics in a harmonically trapped gas. Importantly, in our homogeneous system this implies a {\it global} (quasi-)equilibrium.

Quantitatively understanding the apparently universal shape of the $\Delta N /N$ curves, and their heights, is an interesting  and challenging problem for future work.
Based on the recent observation with $^{85}$Rb~\cite{Klauss:2017}, our atom-molecule mixture likely contains trimers in the first excited Efimov state, which could lead to a nontrivial dependence of the peak value of $\Delta N/N$ on the initial density.
This state has on-resonance size of the order of $1~\mu$m~\cite{Braaten:2007,Zaccanti:2009,Roy:2013,Mestrom:2017} and could set a scale that separates `small' and `large' densities.

In conclusion, we have performed a comprehensive study of the particle-loss, energy, and correlation dynamics in an initially degenerate homogeneous Bose gas quenched to unitarity. We have demonstrated the anticipated scaling laws characterising both a degenerate and a non-degenerate gas, observed universal features of both the dynamical crossover to the thermal regime and of the correlation dynamics, and found that the cloud attains a quasi-equilibrium state while it is still degenerate. In the future, it would be interesting to also study the composition of the molecular cloud created by the magnetic-field sweep~\cite{Klauss:2017}, and distinguish the  two- and three-body correlation dynamics at unitarity.

We acknowledge inspiring discussions with Eric Cornell, Debbie Jin and Francesca Ferlaino, and thank Jean-Loup Ville and Jinyi Zhang  for experimental assistance. This work was supported by the Royal Society, EPSRC [Grants No. EP/N011759/1 and No. EP/P009565/1], ERC (QBox), AFOSR, and ARO. R.L. acknowledges support from the E.U. Marie-Curie program [Grant No. MSCA-IF-2015 704832] and Churchill College, Cambridge. N.N. acknowledges port from Trinity College, Cambridge.

%


\begin{thebibliography}{51}
\bibitem [{Uni({\natexlab{b}})}]{UnitaryFootnote2}%
  \BibitemOpen
  \href@noop{} \bibinfo {note} {{W}e prepare the
  quasi-pure {BEC} in a {$50$~nK} deep trap, but for experiments at unitarity
  we raise the trap depth to {$2~\mu$K}, to avoid any atom loss due to spilling
  or evaporation.}\BibitemShut {Stop}%
\bibitem [{Uni({\natexlab{c}})}]{UnitaryFootnote3}%
  \BibitemOpen
  \href@noop {}\bibinfo {note} {{W}e exponentially ramp
  {$B$} to {$> 10\, \Delta B$} away from {$B_0$}, so the ramp is essentially
  linear in and close to the unitary regime.}\BibitemShut {Stop}%
\end{thebibliography}

\end{document}